\RequirePackage{fix-cm}
\documentclass[twocolumn,epjc3]{svjour3}  
\smartqed  
\RequirePackage{graphicx}
\RequirePackage{aas_macros}
\newcommand{\lowc}[2]{\ensuremath{#1_\mathrm{#2}} }

\newcommand{\aver}[1]{\ensuremath{\langle #1 \rangle} }

\newcommand{\rmd}{\ensuremath{\mathrm{d}}}
\newcommand{\rme}{\ensuremath{\mathrm{e}}}

\newcommand{\rmA}{\ensuremath{\mathrm{A}}}
\newcommand{\rmn}{\ensuremath{\mathrm{n}}}

\newcommand{\rmB}{\ensuremath{\mathrm{B}}}

\newcommand{\neutronsep}{\lowc{S}{n}}
\newcommand{\crosssec}{\ensuremath{\sigma} }
\newcommand{\stellcs}{\ensuremath{\crosssec^*} }
\newcommand{\labcs}{\lowc{\crosssec}{lab}}
\newcommand{\sigmavnostar}{\aver{\crosssec v} }
\newcommand{\sigmav}{\aver{\stellcs v} }
\newcommand{\kb}{\ensuremath{\mathrm{k}_\mathrm{B}}}
\newcommand{\sigmavlab}{\aver{\labcs v} }
\newcommand{\macs}{\aver{\stellcs} }
\newcommand{\labmacs}{\aver{\labcs}}
\newcommand{\photorate}{\ensuremath{\lambda^*} }
\newcommand{\nld}{\ensuremath{\rho_\mathrm{NLD}}}
\newcommand{\nldsep}{\ensuremath{\rho_\mathrm{NLD}(\neutronsep+E)}}

\journalname{Eur. Phys. J. A}
\begin{document}

\title{Stellar neutron capture reactions at low and high temperature}
\subtitle{}


\author{Thomas Rauscher\thanksref{e1,addr1,addr2}}

\thankstext{e1}{ORCID ID: 0000-0002-1266-0642}


\institute{Department of Physics, University of Basel, 4056 Basel, Switzerland \label{addr1}
           \and
           Centre for Astrophysics Research, University of Hertfordshire, Hatfield AL9 10AB, United Kingdom \label{addr2}
}

\date{Received: date / Accepted: date}

\maketitle

\begin{abstract}
The determination of astrophysical reaction rates
requires different approaches depending on the conditions in hydrostatic and explosive
burning.
The focus here is on astrophysical reaction rates
for radiative neutron capture reactions.
Relevant nucleosynthesis processes not only involve the s-process
but also the i-, r- and $\gamma$-processes, which from the nuclear perspective
mainly differ in the relative interaction energies of neutrons and nuclei, and 
in the nuclear level densities of the involved nuclei. Emphasis is put on the difference
between reactions at low and high temperature. Possible complications in the prediction and
measurement of these reaction rates are illustrated and
the connection between theory and experiment is addressed.
\end{abstract}
\section{Introduction}
\label{sec:intro}
Neutron-induced reactions, and specifically neutron captures, play an important role in a number of nucleosynthesis processes.
This is surprising insofar as free neutrons are unstable and therefore a constant neutron supply is required to maintain an appreciable
level of neutrons in an astrophysical plasma. This supply is provided by neutron-releasing particle reactions (s-process, i-process), photon-induced neutron emission
($\gamma$-process) or neutron-rich environments caused by the weak interaction (r-process) \cite{B2FH,Cameron57,1977ApJ...212..149C,1978ApJS...36..285W,Rauscher2}. These processes not only differ in
the neutron sources but also in the temperature achieved in the plasma and in the nuclear level density at the formation
energy of the compound nucleus created in the reaction. This also determines the dominating reaction mechanism and directly impacts
the choice of theory to predict the rate. It also indirectly affects the experimental setup because it determines which
nuclear properties are of interest to be determined experimentally and even whether the astrophysical reaction rate can be
constrained purely experimentally at all, without the invocation of a model. This has to be considered already in the design of an experiment.


\section{Definitions}
\label{sec:defs}
\subsection{Reaction rate}
\label{sec:rate}
The astrophysical reaction rate (number of reactions per time per unit volume of the plasma) for neutron captures A(n,$\gamma$)B is given by \cite{rauintjmodphys,Rauscherb}
\begin{eqnarray}
r_\mathrm{A}^*&=& n_\mathrm{A} n_\mathrm{n}  \sqrt{\frac{8}{\pi \mu}}
\left(\frac{1}{\kb T}\right)^{3/2} \nonumber \\
&&\times \int\limits_0 ^\infty \stellcs\left(E,T\right) E \rme^{-E/(\kb T)} \,\rmd E  \nonumber
\\
&=& n_\mathrm{A} n_\mathrm{n}  \sigmav \quad, \label{eq:ratee}
\end{eqnarray}
where $n_\rmn$ and $n_\rmA$ are the number densities of neutrons and target nuclei, respectively, $E$ is the centre-of-mass
energy, $\mu$ is the reduced mass of neutron and target nuclide, and $T$ is the plasma temperature. The Boltzmann constant is denoted by $\kb$.
The quantity $\sigmav$ is the reaction rate per particle pair (reactivity) under stellar conditions (see
Sec.\ \ref{sec:gscontrib} for a detailed discussion of the impact of thermal plasma effects), which sometimes is denoted by $\left< \sigma v \right>^*$.
It includes the stellar cross section $\stellcs$ of nucleus A for radiative neutron capture. At low $T$ the stellar cross section may
be identical to the laboratory cross section $\labcs$ that, in principle, is directly measureable (unless the cross section is too low or the target nucleus
unavailable for measurements).

An important property of astrophysical reaction rates using the appropriate \textit{stellar} cross section $\stellcs$ is that there is a direct relation
for the rate of the reverse reaction. For B($\gamma$,n)A in a stellar plasma, the astrophysical photodisintegration rate is (see, e.g., \cite{Rauscherb})
\begin{eqnarray}
\photorate = \sigmav \frac{2g_\rmA}{g_\rmB} \frac{G_0^\rmA}{G_0^\rmB} \left( \frac{2\pi\mu\kb T}{h^2}\right)^{3/2} \rme^{-\neutronsep/(\kb T)}\quad. \label{eq:photo}
\end{eqnarray}
The normalized nuclear partition functions $G$ are given by sums over excited states $i$ ($i=0$ specifies the ground state) of the specified nuclide with excitation energy $E_i$ and spin $J_i$,
\begin{equation}
G_0(T)=1+\frac{1}{g_0}\sum\limits_{i>0} \left(2J_i+1\right) \rme^{E_i/(\kb T)}\quad,
\end{equation}
with $g_0=2J_0+1$.
It is to be noted that the connection between Eqs.\ (\ref{eq:ratee}) and (\ref{eq:photo}) only holds when \textit{stellar} cross sections are used.

To obtain the abundance of a nuclide after an elapsed time, one has to consider the difference between all reaction rates creating the nuclide and all reactions destroying it. This leads to a set of coupled
differential equations (e.g., A(n,$\gamma$)B would be among the reactions destroying nuclide A and creating B, and its reverse reaction B($\gamma$,n)A would be among the reactions destroying B and creating A)
that are called a reaction network. Thus, a simple network would be $\rmd Y_\rmA/\rmd t=n_\rmB \lambda^*_\rmB - r_\rmA^*$, $\rmd Y_\rmB/\rmd t=r_\rmA^*-n_\rmB \lambda^*_\rmB$. Integrating the network over time yields the abundances of the included nuclides, e.g., $Y_\rmA$, $Y_\rmB$.  

\subsection{Relevant energy range}
\label{sec:energy}

Although formally the integration limits in Eq.\ (\ref{eq:ratee}) run from zero to infinite energy, most of the contributions to the integral stem from a comparatively narrow energy range. The energy range for the
dominant contributions is given by the convolution of the energy dependence of the cross section and the energy distribution of the neutrons impinging on a nucleus. The latter is given by $E\rme^{-E/(\kb T)}$ and depends
on the plasma temperature $T$. This implies that most neutrons have an energy around $E=\kb T$ and there are (almost) no neutrons at very low and at very high energy.
As neutrons are not affected by the Coulomb force, there is no Coulomb barrier in neutron captures. Rather, the energy dependence of the cross section is
determined by the angular momentum barrier. In principle, in the absence of resonances the cross section is given by a sum over a range of partial waves (s-, p-, d-, ... waves) corresponding to different angular momentum
quantum numbers ($\ell=0$, 1, 2, \dots),
\begin{eqnarray}
\labcs&=&C_{\ell=0}/\sqrt{E}+C_{\ell=1}\sqrt{E}+C_{\ell=2}E^{3/2}+\dots \nonumber\\
&=&\sum_\ell C_\ell E^{\ell-1/2}\quad.\label{eq:partial}
\end{eqnarray}
At the comparatively low interaction energies encountered in astrophysical environments only few partial waves contribute and the allowed
$\ell$ ($C_\ell >0$) are selected by spin and parity selection rules. The $C_\ell$ depend on nuclear and quantum constants but also on the strength of electromagnetic transitions. The latter depend
on the energy of the released $\gamma$-rays $E_\gamma \leq E+\neutronsep$ and this gives rise to an energy dependence of the $C_\ell$. As long as the interaction energy $E$, however, is small compared to the neutron separation energy \neutronsep
in the final nucleus, the energy dependence is negligible. Along stability $E\ll\neutronsep$ is always fulfilled. Approaching the neutron dripline, this is not the case anymore.

Even with energy-dependent $C_\ell$, in the absence of resonances the energy dependence of the cross section is weaker than the energy dependence of the neutron energy distribution. Therefore this determines
the energy range of the main contributions to the integral in Eq.\ (\ref{eq:ratee}) is only slightly modified. A good approximation is \cite{1969ApJS...18..247W,rauscher3}
\begin{eqnarray}
E_\mathrm{eff}\approx 1.72\times10^{-10} T \left(\ell+1/2\right)\quad\mathrm{MeV}\,, \label{eq:gamow}\\
\Delta_\mathrm{E} \approx 1.94\times10^{-10} T \sqrt{\ell+1/2}\quad\mathrm{MeV}\,,
\end{eqnarray}
with the energy range given by $E_\mathrm{eff}\pm\Delta_\mathrm{E}$ when the temperature is given in K. Although the dominant $\ell$ may not always be known, the shifts with increasing $\ell$ are small and the values for $\ell=0$
provide a reasonable guidance \cite{rauscher3}. In fact, even for $\ell=0$ the value of $E_\mathrm{eff}$ is close to $\kb T$. Again, these relations are applicable for most nuclides and only lose their validity close to the neutron dripline
for nuclides with small \neutronsep, as explicitly shown by \cite{rauscher3}. Strictly speaking, the notion of a single energy range contributing to the rate integral is only valid for smooth, non-resonant cross sections. In the presence of
$i$ isolated
resonances with their resonance energies $E_\mathrm{r}^i$, 
the rate can be described by a (coherent) sum of resonance contribution with their individual effective energy ranges. This could also be incorporated as (strongly) energy dependent coefficients $C_\ell^i (E,E_\mathrm{r}^i)$.
In the regime of unresolved, overlapping resonances the situation reverts to the non-resonant case with a single range of effective energies. In short, assuming cross sections at energies around $\kb T$ are the dominant contributors to
the reaction rate integral shown in Eq.\ (\ref{eq:ratee}) is safe for most astrophysical reactions.

\subsection{MACS}
\label{sec:macs}

Experimental investigations of neutron captures often report the \textit{Maxwellian Averaged Cross Section} (MACS) \labmacs instead of the capture cross section $\labcs$, defined as (e.g., \cite{Rauscherb})
\begin{eqnarray}
\labmacs &=& \frac{2}{\sqrt{\pi}}
\frac{1}{\left(\kb T\right)^2} \int_0 ^\infty \labcs(E) E \rme^{-E/(\kb T)} \,\rmd E  \nonumber
\\
&=& \sigmavlab/v_\mathrm{T} \quad, \label{eq:labmacs}
\end{eqnarray}
with the thermal (most probable) velocity 
\begin{equation}
v_T=\sqrt{(2\kb T)/\mu}\quad.
\end{equation}
An interesting relation in the context of the MACS is the fact that $\labmacs v_\mathrm{T}=\sigmavlab=\mathrm{const}$ for s-wave neutron capture.

Historically, there is a theoretical and an experimental motivation for introducing the MACS. On the theory side, in the \textit{classical s-process} model a full reaction network was simplified by only considering neutron capture reactions (neglecting the reverse, photodisintegration
reactions because they are too slow at regular s-process conditions) and $\beta^-$ decays, and by assuming steady flow, i.e. the abundances have reached their equilibrium values and don't change over time  \cite{1961AnPhy..12..331C}. Further assuming that s-wave neutron capture
dominates the (n,$\gamma$) cross sections and using a neutron flux $n_\rmn v_\mathrm{T}$, it can be shown (e.g., \cite{Rauscherb}) that in the (local) steady-flow equilibrium there is a connection between the \textit{stellar} MACS and the
abundances of neighboring nuclides,
\begin{equation}
\macs_\rmA Y_\rmA = \macs_\rmB Y_\rmB = \mathrm{const}\quad.
\end{equation}
The constant is determined by the actual neutron exposure. Using long-term neutron exposures with an exponential decay in time, the classical s-process model identified several s-process contributions to the abundances, the main and weak s-process
(a third component, the strong s-process, was also discussed for a while) \cite{1965ApJS...11..121S}. The MACS is taken at $\kb T=30$ keV, a typical value for He-shell flashes in AGB stars.

On the experimental side, Beer and K\"appeler \cite{1979LIACo..22...79H,1980PhRvC..21..534B} pioneered a method to directly measure the MACS by activation using a tailored neutron spectrum (obtained from the $^7$Li(p,n)$^7$B reaction) corresponding to a thermal energy
spectrum at $\kb T=25$ keV, very close to the energy distribution assumed for AGB stars. This not only allowed to directly determine the quantity required in the classical model but also solved experimental complications with time-of-flight (TOF) measurements
and with the definition of the neutron spectrum for the activation method \cite{1982ApJ...257..821K,2014JPhG...41e3101R}. This ground-breaking approach led to a wealth of experimental MACS data, also made available in dedicated compilations \cite{2000ADNDT..76...70B,2006AIPC..819..123D,2014NDS...120..171D}
for the s-process and also boosted theoretical s-process studies. Not all reactions of interest can be measured by activation, though, as this method requires an unstable final nucleus B. Therefore the activation measurements are supplemented by high-resolution
TOF measurements probing the energy range required to compute the MACS. Unfortunately, often such measurements only published the derived 30 keV MACS instead of the measured cross sections. This loses information
that could be helpful for a further theoretical analysis, especially when rates at other temperatures are needed.

With the advances in stellar models combined with precise nuclear and astronomical data, it has become apparent that the classical s-process model is not sufficient to explain all features of s-process nucleosynthesis \cite{1999ApJ...525..886A}.
It also became clear that AGB stars are not the only site of the s-process but that also massive stars contribute. Even within AGB stars, one has to consider two production regions with different timescales and temperatures. Therefore, modern
models have to abolish the simplifying assumptions of steady flow and a single temperature and use complete reaction networks requiring the knowledge of reaction rates across a larger range of temperatures, extending well below and well above 30 keV. This necessitates to go beyond a 30 keV MACS (or an experimental MACS at another
discrete temperature, as could be achieved by using another reaction for neutron production \cite{2014JPhG...41e3101R}). Such reaction rates are either derived from experimentally measured cross sections across a sufficiently large energy range,
from theoretical predictions, or from a combination of both. Publicly available reaction network codes, inspired by high-$T$ nucleosynthesis and suited for a large range of temperatures, make use of reaction rates as defined in Eq.\ \ref{eq:ratee} instead of MACS.
Such rates are compiled as tables or fits. This eventually leads to a phase-out of MACS in models.

Another reason why the MACS loses its importance in modern astrophysical investigations is the fact that laboratory measurements cannot directly measure the astrophysical rate at higher temperature because of the increasing
contributions from excited target states (see Sec.\ \ref{sec:gscontrib}). Thus, a laboratory measurement of the MACS $\labmacs$ does not yield immediately the astrophysically interesting quantities $\macs$ or $\sigmav$ and the original experimental advantage and motivation to
determine the MACS is lost even when the activation method is applicable. Even when the MACS or rate cannot be constrained fully by experiment alone, high-resolution cross section measurements across a range of energies
nevertheless can help to test nuclear theory and its predictions of cross sections or other nuclear properties. A conversion of these data to MACS, however, becomes unnecessary.

\subsection{Ground-state contribution to the stellar rate}
\label{sec:gscontrib}

\begin{figure}
  \includegraphics[width=\columnwidth]{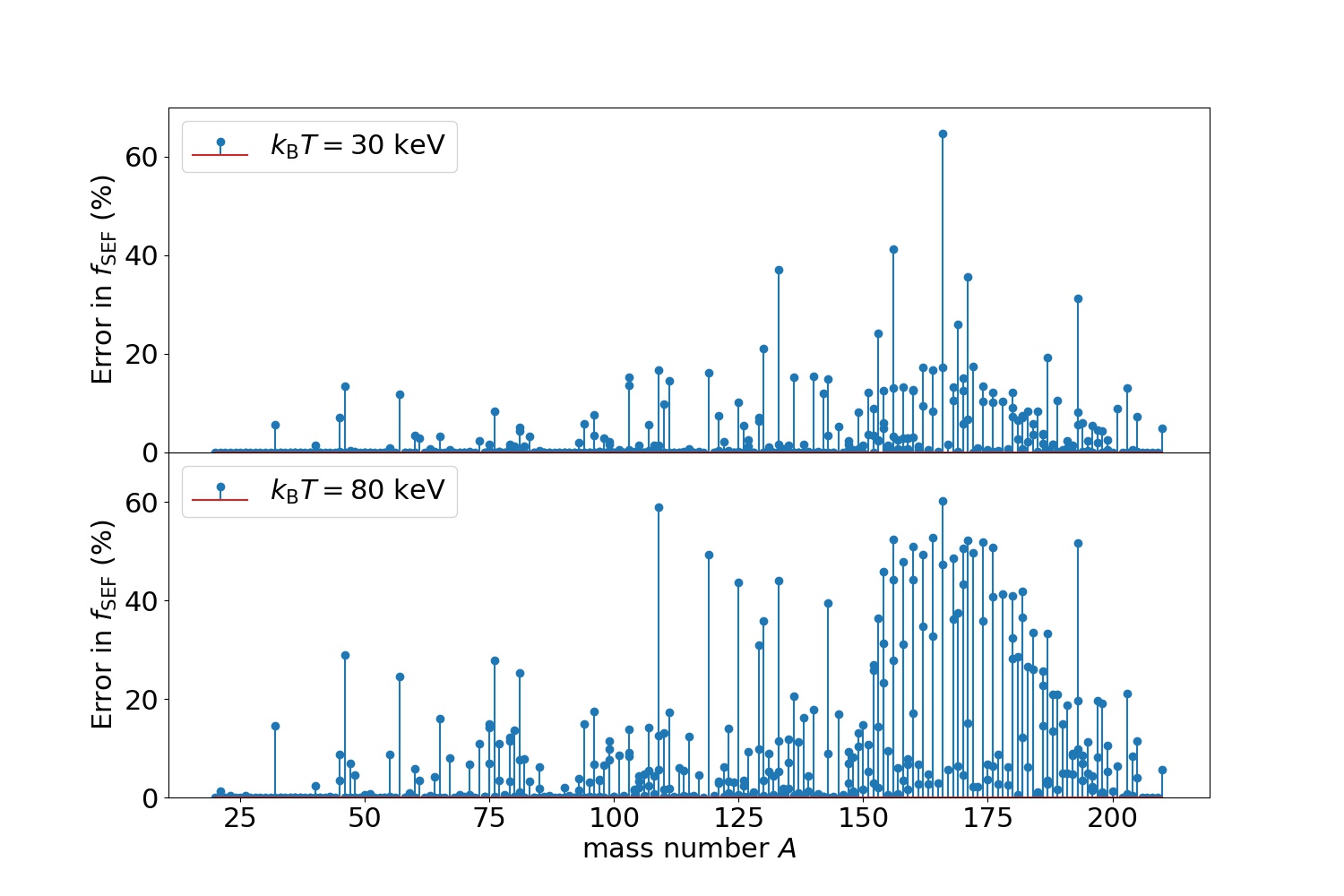}
\caption{Percentage error in SEF estimate of the thermal excited state contribution.}
\label{fig:seferror}       
\end{figure}

\begin{table}
\caption{Nuclides with $X_0^*\leq 0.8$ at $\kb T=30$ keV.}
\label{tab:excited}       
\begin{tabular}{rrrrrr}
\hline\noalign{\smallskip}
 $^{57}$Fe &  $^{73}$Ge &  $^{80}$Br &  $^{83}$Kr &  $^{94}$Nb &  $^{96}$Tc   \\
$^{103}$Ru & $^{105}$Ru & $^{103}$Rh & $^{104}$Rh & $^{107}$Ag & $^{108}$Ag   \\
$^{111}$Ag & $^{119}$Sn & $^{121}$Sn & $^{121}$Sb & $^{122}$Sb & $^{126}$Sb   \\
$^{127}$Te &  $^{130}$I & $^{129}$Xe & $^{134}$Cs & $^{140}$La & $^{133}$Ce   \\
$^{142}$Pr & $^{151}$Sm & $^{153}$Sm & $^{154}$Sm & $^{151}$Eu & $^{152}$Eu   \\
$^{156}$Eu & $^{153}$Gd & $^{155}$Gd & $^{156}$Gd & $^{157}$Gd & $^{158}$Gd   \\
$^{160}$Gd & $^{158}$Tb & $^{159}$Tb & $^{160}$Tb & $^{161}$Tb & $^{159}$Dy   \\
$^{161}$Dy & $^{162}$Dy & $^{164}$Dy & $^{166}$Ho & $^{162}$Er & $^{164}$Er   \\
$^{168}$Er & $^{169}$Er & $^{170}$Er & $^{169}$Tm & $^{170}$Tm & $^{171}$Tm   \\
$^{170}$Yb & $^{171}$Yb & $^{172}$Yb & $^{174}$Yb & $^{176}$Yb & $^{174}$Hf   \\
$^{178}$Hf & $^{180}$Hf & $^{181}$Hf & $^{182}$Hf & $^{179}$Ta & $^{180}$Ta   \\
$^{182}$Ta &  $^{182}$W &  $^{183}$W &  $^{185}$W & $^{186}$Re & $^{188}$Re   \\
$^{189}$Os & $^{192}$Ir & $^{193}$Ir & $^{194}$Ir & $^{193}$Pt & $^{197}$Pt   \\
$^{201}$Hg &  &  &  &  &  \\
\noalign{\smallskip}\hline
\end{tabular}
\end{table}

\begin{figure*}
  \includegraphics[width=\textwidth]{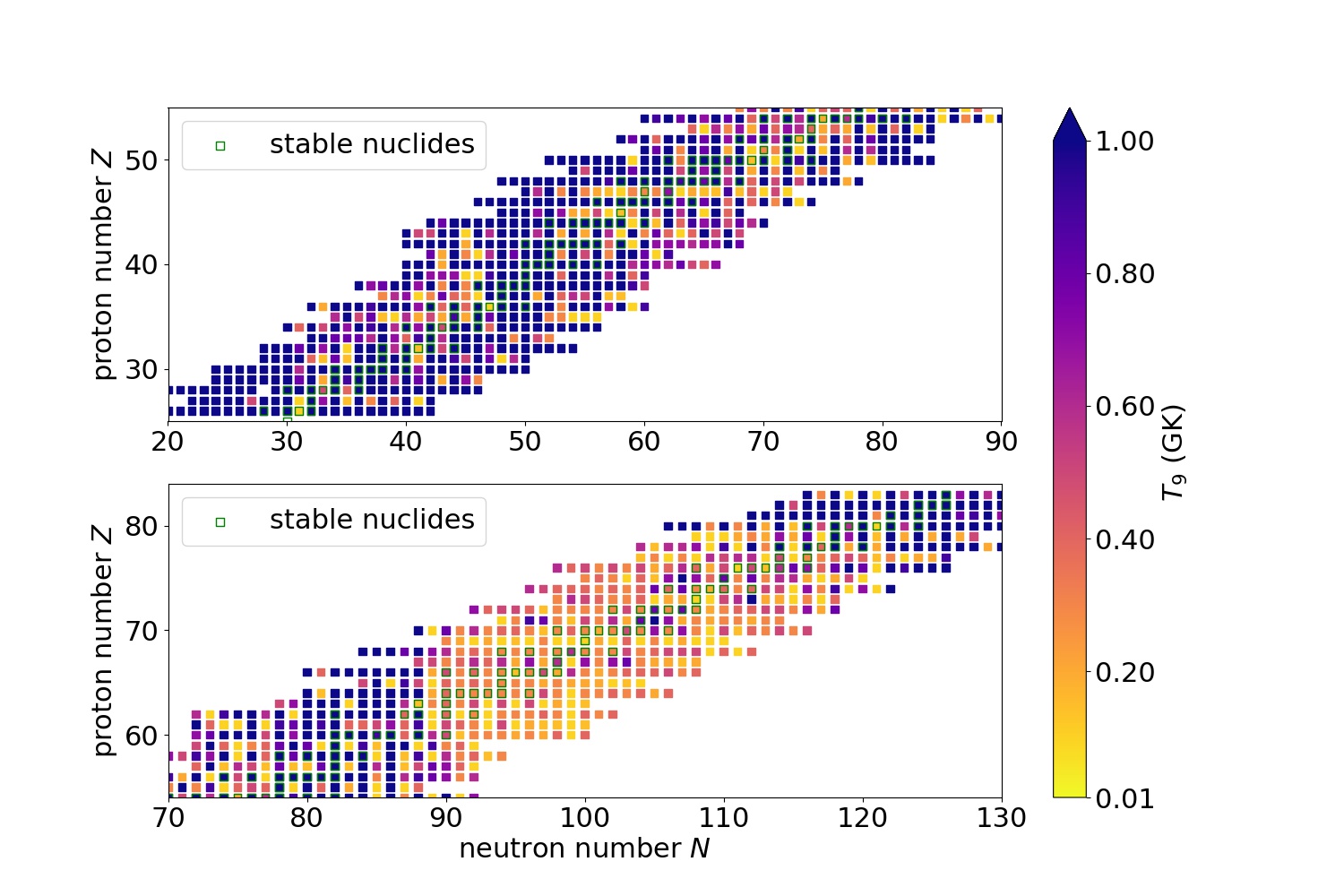}
\caption{Temperature at which $X_0^*\leq0.8$.}
\label{fig:xtemp}       
\end{figure*}

Often misjudged is the impact of thermal modifications of the reaction cross sections in a stellar plasma. Nuclei in such a plasma can be excited through thermal and nuclear interactions and therefore a fraction of the target nuclei is found
in an excited state \cite{1980ApJ...238..266W}. The population of an excited state $i$ in a nucleus at a given plasma temperature $T$ is given by the Boltzmann factor 
\begin{equation}
B_i(T)=g_i \rme^{-E_i/(\kb T)}\label{eq:boltz}
\end{equation}
and the population relative to the ground
state is
\begin{equation}
\mathcal{P}_i = \frac{g_i}{g_0 G_0} \rme^{-E_i/(\kb T)}=\frac{B_i}{(2J_0+1)G_0}\quad.\label{eq:pop}
\end{equation}
To derive the stellar cross section $\stellcs$ as used in Eq.\ (\ref{eq:ratee}) one has to realize that each nucleus -- in the ground state or in an excited state
-- is bombarded by neutrons with an energy distribution given by the plasma temperature. This means that the rate including nuclei in thermally excited states comprises a sum of rates, one for the ground state and each excited state,
\begin{eqnarray}
\sigmav &=& \mathcal{P}_0 \sigmavnostar_0 + \mathcal{P}_1 \sigmavnostar_1 + \mathcal{P}_2 \sigmavnostar_2 + \dots \nonumber \\
&=& \sum\limits_i \mathcal{P}_i \sigmavnostar_i \nonumber\\
&=& \sqrt{\frac{8}{\pi\mu}} \left(\frac{1}{\kb T}\right)^{3/2} \nonumber \\
&&\times \sum\limits_i \mathcal{P}_i \int\limits_0^\infty \sigma_i(\epsilon_i) \epsilon_i \rme^{-\epsilon_i/(\kb T)} \rmd \epsilon_i\,. 
\end{eqnarray}
The cross section of the nucleus in excited state $i$ is denoted by $\sigma_i$. Note that the integration variable $\epsilon_i$ runs from zero to infinity in each case because each excited state is bombarded by the same neutron
distribution. Relative to each other, however, the integrals are shifted by the excitation energies $E_i$. In order to arrive at Eq.\ (\ref{eq:ratee}) with a single integral instead of a weighted sum of integrals, it is necessary
to realize that it is mathematically permitted to exchange summation and integration, as pointed out by \cite{1974QJRAS..15...82F}. To collapse the range of integrals to a single integral, their energy scales $\epsilon_i$
have to be shifted by the $E_i$ and the integration limits adjusted accordingly. The full derivation is given in \cite{Rauscherb}. Comparing the result to Eq.\ (\ref{eq:ratee}) allows to find the expression for the stellar
cross section,
\begin{eqnarray}
\stellcs (E,T) &=& \frac{1}{\left(2J_0+1\right) G_0^\rmA(T)} \nonumber \\
&&\times\sum\limits_i \left(2J_i+1\right) \left(1-\frac{E_i}{E}\right) \sigma_i (E-E_i)\quad.\label{eq:stellcs}
\end{eqnarray}
The individual cross sections $\sigma_i$ for reactions on nuclei in the $i$-th excited state now have to be evaluated at an energy $E-E_i$. Following \cite{1974QJRAS..15...82F}, cross sections at $(E-E_i) \leq 0$ are set to zero.
A temperature dependence of the stellar cross section enters through the $T$-dependence of the normalized partition function $G_0^\rmA(T)$ of the target nuclide A. At low $T$ the peak of the neutron energy distribution (Eq.\ \ref{eq:gamow})
is shifted well below the excitation energy of the first excited state $E_1$. Then only reactions on the nuclear ground state contribute significantly because most neutrons do not have sufficient energy to allow for non-zero $\sigma_{i>0}$.
This situation is (almost) equivalent to using the laboratory cross section $\labcs$ instead of $\stellcs$ in Eq.\ (\ref{eq:ratee}).
With rising $T$ the number of neutrons at higher energy increases and more and more $\sigma_i$ of excited states provide non-negligible contributions. The size of the individual contributions depends on the actual energy dependences of the cross sections $\sigma_i$
and on the weighting factor $W_i$ appearing in front of the $\sigma_i$ in Eq.\ (\ref{eq:stellcs}), with
\begin{equation}
W_i(E)=\left(2J_i+1\right) \left(1-\frac{E_i}{E}\right)\quad.
\end{equation}
Interestingly, $W_i$ shows a linear dependence on the energy of the excited state $E_i$ whereas the Boltzmann factor $B_i$ falls off exponentially with $E_i$. The actual contribution, however, is difficult to assess from $W_i$ alone because
a range of c.m.\ energies $E$ is contributing to the rate integral (see Sec.\ \ref{sec:energy}). Moreover, the behaviour of the $\sigma_i$ at low energy significantly impacts the relative importance of a contribution. This is especially relevant in
s-wave neutron captures because the cross section increases with $1/\sqrt{E-E_i}$ towards small $E-E_i$. Therefore contributions from excited states are expected to be more important in neutron captures dominated by s-waves on excited states.
This is the dominant partial wave in the majority of neutron captures. It was found that 
excited states up to $E_i\approx \kb T$ may contribute significantly \cite{rauintjmodphys,Rauscherb}.

\begin{figure*}
  \includegraphics[width=\textwidth]{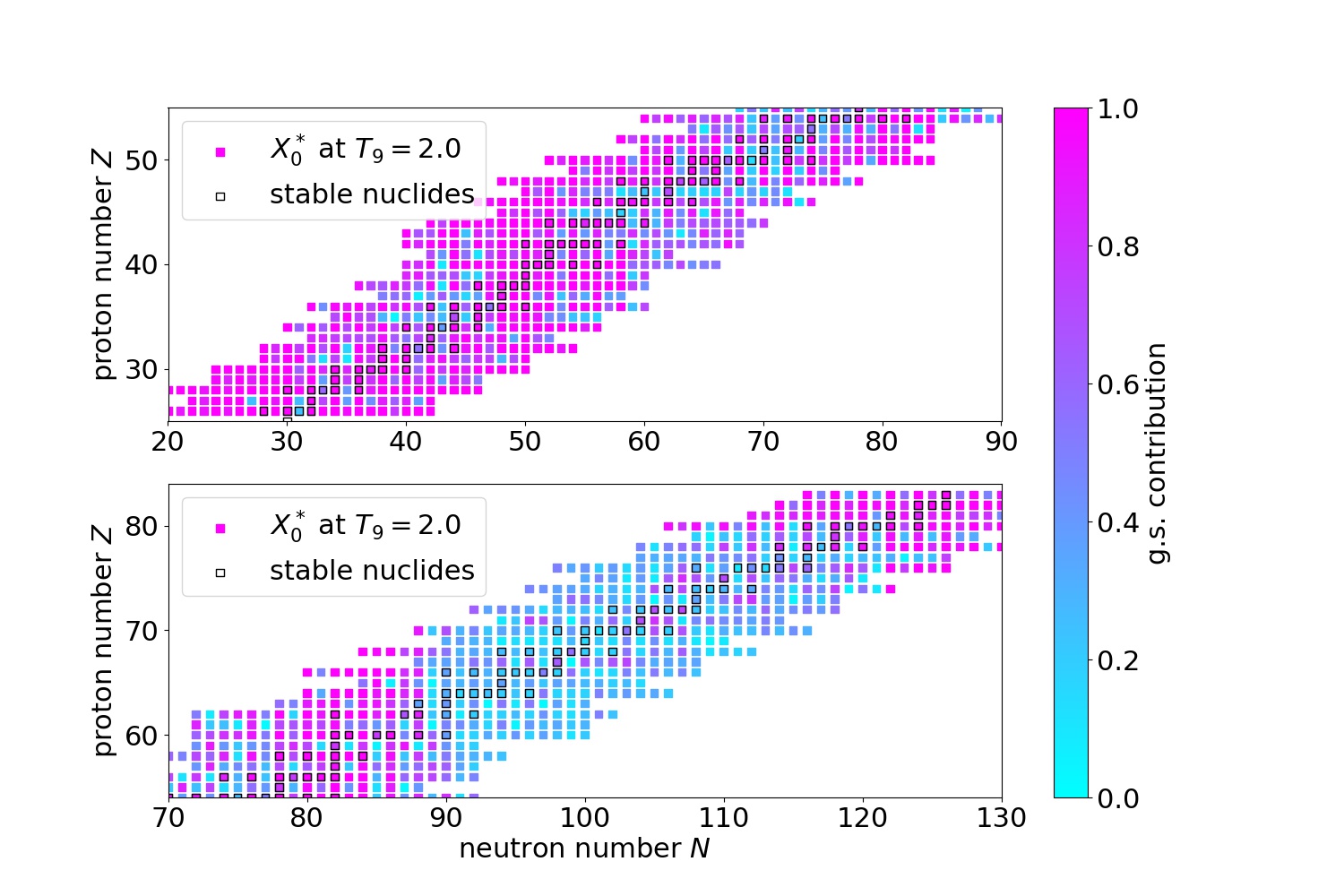}
\caption{Ground-state contribution $X_0^*$ at $T=2$ GK.}
\label{fig:x2}       
\end{figure*}

From Eq.\ (\ref{eq:pop})  it is easy to see that the actual contribution $X_i^*$ of reactions on level $i$ to the astrophysical reactivity
is
\begin{equation}
X_i^* (T)= \frac{P_i \sigmavnostar_i}{\sigmav}=\frac{g_i}{g_0 G_0}\rme^{-E_i/(\kb T)}\frac{\sigmavnostar_i}{\sigmav}\quad.
\end{equation}
For the ground-state (g.s.) contribution $X_0^*$ this reduces to
\begin{equation}
X_0^*(T)=\frac{1}{G_0}\frac{\sigmavnostar_0}{\sigmav}=\frac{1}{G_0}\frac{\sigmavnostar_\mathrm{lab}}{\sigmav}\,.
\end{equation}
The g.s.\ contribution is a monotonically dropping function with increasing $T$, in the range $0\leq X_0^*(T)\leq 1=X_0^*(0)$.
The combined contribution of all excited states to the astrophysical rate simply is $X_\mathrm{exc}^*=1-X_0^*$.

It is important to realize that this is different from a simple comparison of astrophysical and laboratory reactivities (or rates) as it is often found in literature when using the stellar enhancement factor (SEF) $f_\mathrm{SEF}=\sigmav/\sigmavnostar_\mathrm{lab}$. 
The SEF is not a measure of the importance of excited state contributions because it does not account for the population of excited states through the partition function $G_0$.
Thus, it overestimates the g.s.\ contribution at non-zero temperature and underestimates the relative contribution of excited states.
A value $f_\mathrm{SEF}\simeq1$ does \textit{not} support the conclusion
that the excited states do not contribute nor that the rate is fully constrained by a determination -- for example by a measurement -- of $\sigmavnostar_\mathrm{lab}$ \cite{2011ApJ...738..143R}. The combined contributions of reactions on ground and excited states could just
sum up to yield $\sigmav\simeq\sigmavnostar_\mathrm{lab}$ but with non-negligible $X_\mathrm{exc}^*$. For the same reason a rescaling of an experimentally determined $\sigmavnostar_\mathrm{lab}$ by $f_\mathrm{SEF}$ to obtain $\sigmav$ is highly
questionable. Figure \ref{fig:seferror} shows the (underestimation) error in the SEF for stable nuclides (see also \cite{2011ApJ...738..143R} for a list
of these nuclides along with their values of $X_0^*$ and $f_\mathrm{SEF}$) due to the neglect of the partition function.
It basically reproduces the values of $G_0$. These are well known around stability because the excitation energies and spins
of the contributing, low-lying levels are experimentally determined. Table \ref{tab:excited} lists naturally occurring nuclides
up to Bi that exhibit $X_0^*\leq 0.8$ already at $\kb T=30$ keV. Neutron capture rates on these nuclides cannot be constrained
accurately by a measurement without invoking additional theoretical considerations (see \cite{2012ApJ...755L..10R} for a
detailed discussion of how to combine theory and experiment in such cases).
Figure \ref{fig:xtemp} shows the temperature at which the g.s.\ contribution drops to 80\% and below for nuclides at and
around stability (data taken from \cite{2012ApJS..201...26R}). It can be clearly seen that the higher intrinsic nuclear level density of intermediate and heavy nuclides -- and
in particular in strongly deformed nuclei -- reduces the g.s.\ contribution already at low plasma temperature. There are some
exceptions in the lighter nuclides, for which the g.s.\ contribution is low already at s-process temperatures (see also Table \ref{tab:excited}).
Most notable, for example, is $^{57}$Fe with a first excited state at 14.4 keV, leading to $X_0^*=0.39$ at $\kb T=30$ keV.
Nevertheless, the SEF is only $f_\mathrm{SEF}=1.1$, which would incorrectly suggest only a small contribution of excited states.

For comparison, the temperature in the s-process ranges from
$\kb T=8$ keV ($T=0.09$ GK; AGB interpulse burning), over $\kb T=22$ keV ($T=0.25$ GK; convective He-core burning of massive stars)
and $\kb T=30$ keV ($T=0.384$ GK; thermal AGB pulses), to $\kb T=90$ keV ($T=1.04$ GK; C-shell burning in massive stars). 
Inspecting Table \ref{tab:excited} and Fig.\ \ref{fig:xtemp} it becomes evident
that already at temperatures corresponding to the thermal pulses of AGB stars, contributing to the main s-process component,
the g.s.\ contributions for many nuclides are already small and that for temperatures of C-shell burning in massive stars,
contributing to the weak s-process component, excited state contributions dominate the reaction rate for almost all nuclides
in the s-process path along stability. This has important consequences for experiments because it does not allow to directly constrain
the astrophysical reaction rate by measuring capture cross sections or the MACS of nuclei in their ground states.

The fact that the contributions of thermally excited states are important in astrophysical reaction rates has been well
established in the community studying explosive nucleosynthesis at high $T$. At temperatures of a few GK, all rates are
dominated by these contributions and the g.s.\ contribution becomes small or even negligible. Figure \ref{fig:x2} illustrates
this by plotting $X_0^*(\kb T=2\mathrm{GK})$ for nuclides at and around stability.

\section{Further differences between neutron captures at low and high temperature}
\label{sec:theory}

Beyond the s-process other nucleosynthesis processes involving neutrons are the i-, r-, and $\gamma$-process. While the i-process
involves temperatures comparable to core He-burning in massive stars, the r-process already exceeds this temperature, proceeding at
$1-2$ GK. The $\gamma$-process in the outer shell of an exploding massive star photodisintegrates intermediate and heavy nuclides
with emission of neutrons and charged particles at $2-3.5$ GK. The emitted neutrons can then be recaptured by other nuclides.

\subsection{Reverse rates}
\label{sec:reverse}

The magnitude of excited state contributions is not the only difference between nucleosynthesis at low and at high temperature.
In order to follow the abundance evolution in high-$T$ environments it becomes necessary to also include the reverse reactions
into the network. This is easily seen in Eq.\ (\ref{eq:photo}) because the ratio of reverse to forward rate is proportional to 
$\exp\left(-\neutronsep/(\kb T)\right)$. This explains why a simpler network only containing neutron captures and $\beta^-$ decays
along the line of stability is sufficient for s-process simulations whereas also ($\gamma$,n) reactions have to be included
for the i-, r-, and $\gamma$-processes. Excited state contributions to the reaction rate play a dominant role in these circumstances.
It is interesting to note, however, that captures still have larger $X_0^*$ by several orders of magnitude than ($\gamma$,n) reactions 
especially at $\gamma$-process temperatures and that this makes an experimental determination of the capture rate highly preferrable 
over a photodisintegration measurement \cite{2008PhRvL.101s1101K,2009PhRvC..80c5801R}.

With sufficiently high neutron densities $n_\rmn$, as
attained in the i- and r-processes, forward and reverse rate become comparably fast and reach an (n,$\gamma$)-($\gamma$,n) equilibrium.
The individual nuclide abundances in such an equilibrium do not depend on the rates anymore (provided the rates stay fast enough
to remain in equilibrium) and can be calculated from simpler relations derived by equating
Eqs.\ (\ref{eq:ratee}) and (\ref{eq:photo}), wherein $\sigmav$ cancels out \cite{Rauscherb}. This implies that the 
(experimental or theoretical) knowledge of cross sections and rates is only required to follow the freeze-out from
equilibrium with dropping temperature. Depending on the process and the astrophysical simulation, the freeze-out can be so fast that
final neutron captures do not alter the equilibrium abundances significantly \cite{2005NuPhA.758..655R,2014JPhG...41e3101R}.

\subsection{Systems with low, intermediate, and high intrinsic nuclear level density}
\label{sec:nld}

The high neutron densities in the i- and r-process together with the fact that reaction rates depend exponentially
on temperature, allows for the (temporary) production of highly unstable nuclides in these high-$T$ environments. These require different
experimental approaches than used for stable species. Then it is especially important to understand what kind of information is
helpful to improve theoretical models, also because high-$T$ reaction rates cannot be constrained by measurements on nuclei in their ground states.

All the high-$T$ processes mostly involve intermediate and heavy nuclides. These may, however, have been built by reactions on light nuclides.
Also in the s-process light nuclides play a role because they can act as neutron "poisons" when exhibiting a large neutron capture cross section or when being
very abundant in the plasma. Neutron poisons remove neutron flux and thus hamper the production of heavier nuclides through neutron captures.
Regardless of the mass of the nucleus, the actual distinguishing
feature is the nuclear level density (NLD) $\nld$ at the compound formation energy, i.e., the (hypothetical) excitation energy at which
the nucleus B=A+n is formed. The compound formation energy for neutron captures is given by the sum of neutron separation energy in B, $\neutronsep$, and 
the c.m.\ energy $E$.
Thus,
the interesting quantity is $\nldsep$ in nucleus B. (The neutron energy $E$ is negligible compared to $\neutronsep$ for basically all astrophysical neutron captures except very close to the neutron dripline.)
Different approaches have to be adopted for experimental and theoretical studies of systems with low, intermediate, and high NLD.

It is important to consider that different reaction mechanisms are dominating the captures in these systems. Light nuclides exhibit large level spacings
and therefore $\nldsep$ is small, despite of large $\neutronsep$. 
Without levels close to $\neutronsep$, the \textit{direct reaction mechanism} dominates, directly
capturing the neutron into its final state in nucleus B and radiating away the excess energy as a mono-energetic $\gamma$-ray \cite{rauintjmodphys,Rauscherb}. This process can be described in
a potential model, using effective nuclear potentials to calculate the wavefunctions of incident and captured neutron, and a simple multipole operator
to account for the electromagnetic emission. Depending on the dominating partial wave, the obtained cross sections show a behaviour as given by Eq.\ (\ref{eq:partial}).
Very light systems (typically $A\leq20$) can also be described in more microscopically grounded models, making use of effective nucleon-nucleon interactions (see, e.g.,
\cite{2006NuPhA.777..137D} for references).

With increasing NLD around $\neutronsep$, neutrons can be captured through resonances or tails of resonances into excited nuclear states, sharing their initial energy
among all the nucleons in the system. This is the \textit{compound reaction mechanism}, forming an excited compound nucleus that subsequently decays through $\gamma$-cascades
(and particle emission, if energetically possible). This is modeled by (partial) resonance widths (related to transmission coefficients) derived from particle wavefunctions in effective nuclear potentials, usually
applying the optical model. The $\gamma$-width is computed by folding a $\gamma$-strength function (specifying the probability for the emission of the specific EM radiation between two states) with the number
of available final states at an excitation energy given by $\neutronsep+E-E_\gamma$, where $E_\gamma$ is the energy of the $\gamma$-ray. A reliable prediction of resonant cross section is
difficult due to interference effects between resonances and between a resonance and the direct capture background. Often, phenomenological approaches are used, such as the R-matrix method which
fits resonance properties to measured excitation functions. An independent determination of resonance widths is helpful to improve predictions for nuclides for which cross section measurements are
unavailable.

\begin{figure}
  \includegraphics[width=\columnwidth]{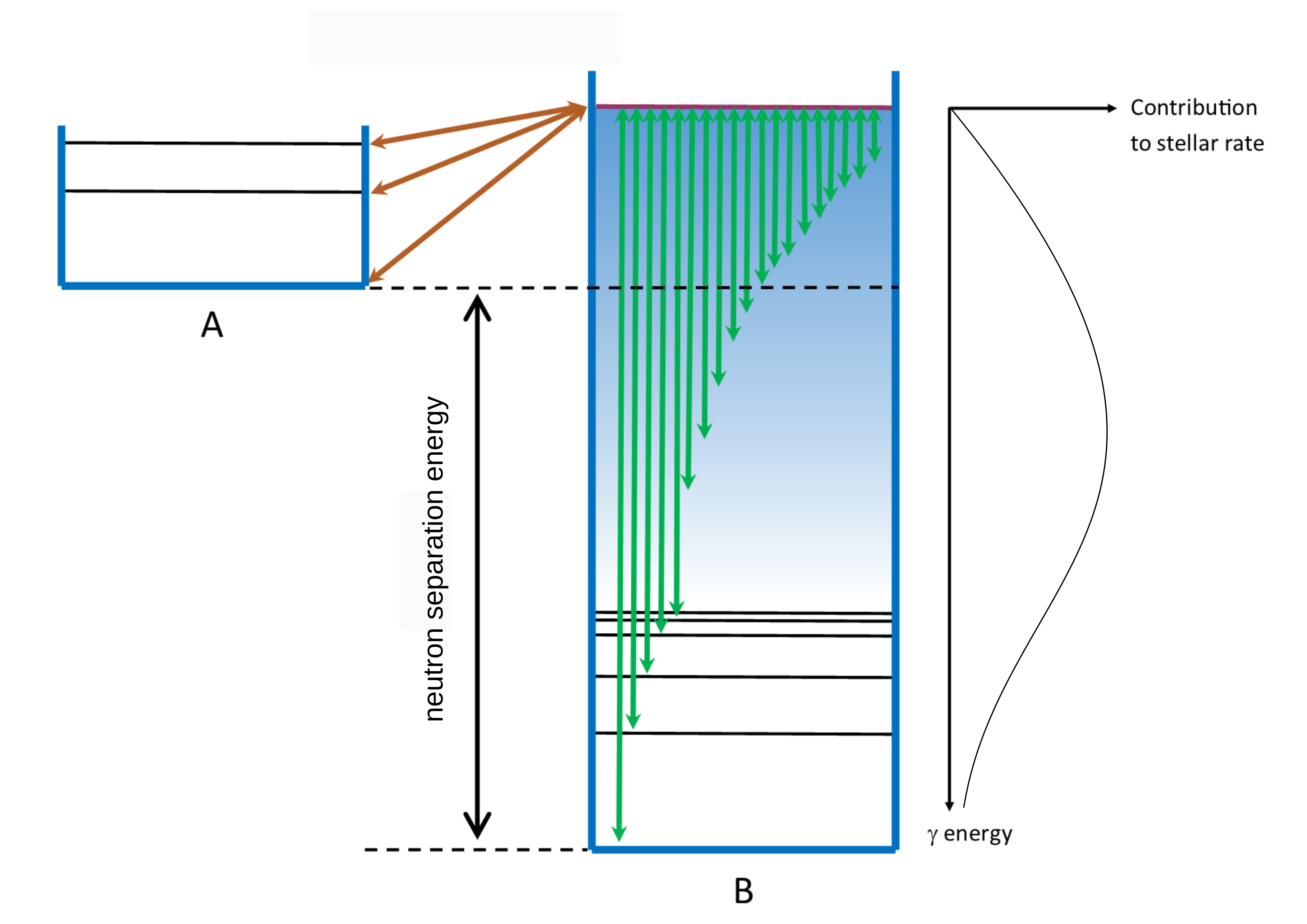}
\caption{Schematics of the relative importance of $\gamma$-energies contributing to the capture cross section and reaction rate, not to scale. [Figure by the author, first published in \cite{2018arXiv180310581R}.]}
\label{fig:gamma}       
\end{figure}

\begin{figure}
  \includegraphics[width=\columnwidth]{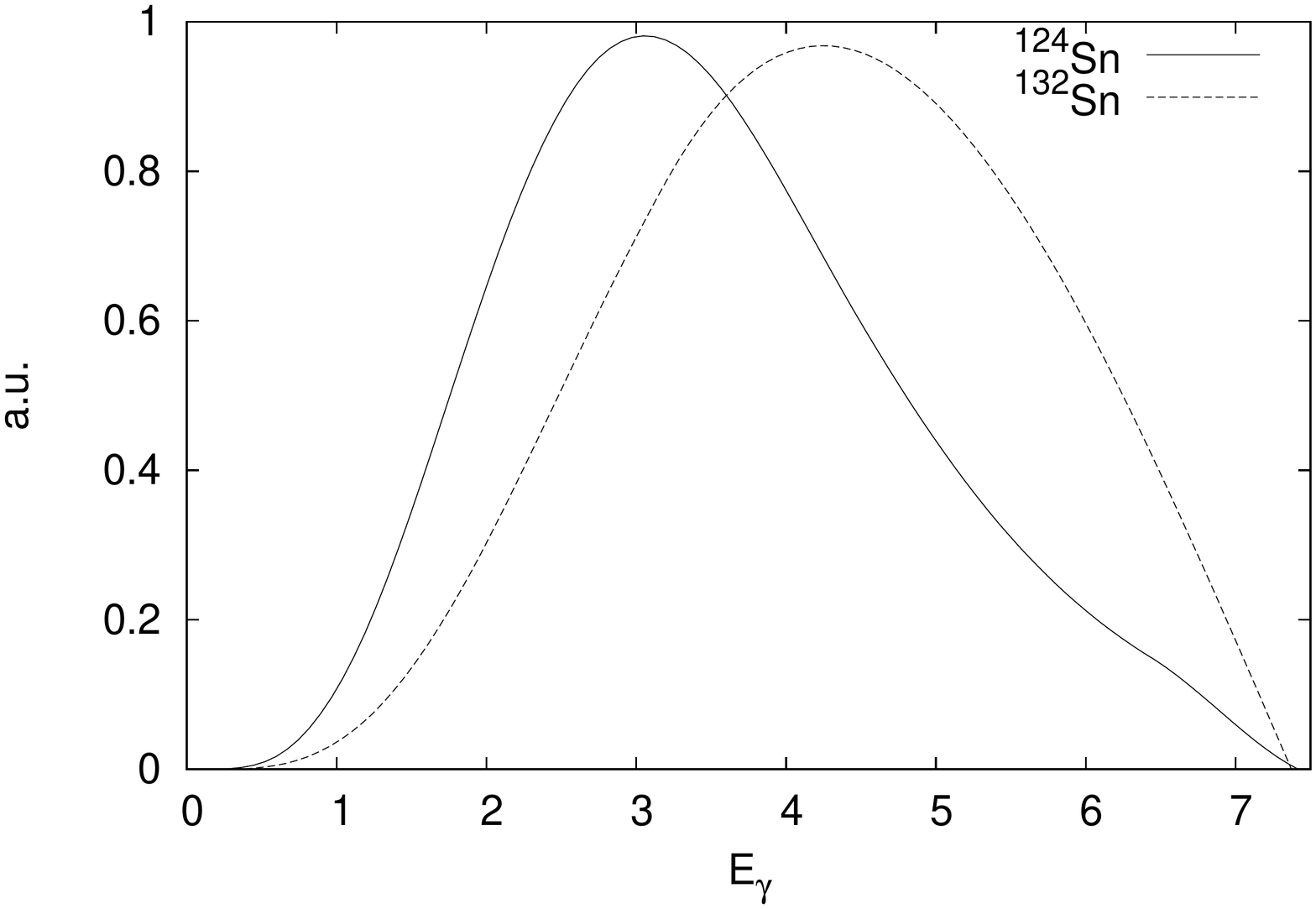}
\caption{Relative contribution of $E_\gamma$ to the reaction rate for the capture of 60 keV neutrons on two Sn isotopes. [Figure from \cite{2008PhRvC..78c2801R}, with permission.]}
\label{fig:maxcontrib}       
\end{figure}

The NLD is very high around the compound formation energy in intermediate and heavy nuclides, which comprise the majority of nuclei in neutron-capture nucleosynthesis. The compound reaction mechanism is definitely
dominating in this case, with the direct mechanism being negligible. At high NLD individual resonances cannot be disentangled anymore and this feature lends itself to apply a model using average properties, such
as average widths. This is called the \textit{statistical model of compound nuclear reactions} or the \textit{Hauser-Feshbach model} \cite{rauintjmodphys,Rauscherb}. It assumes the presence of resonances with any spin and
parity at the compound formation energy. Instead of the widths of the individual resonances,
averaged particle widths for each spin/parity are calculated from optical model potentials. For the $\gamma$-width, the averaged property again is based on the $\gamma$-strength function, with E1 transitions dominating in most cases but also M1 and E2
transitions can be considered.

\begin{figure*}
  \includegraphics[width=\textwidth]{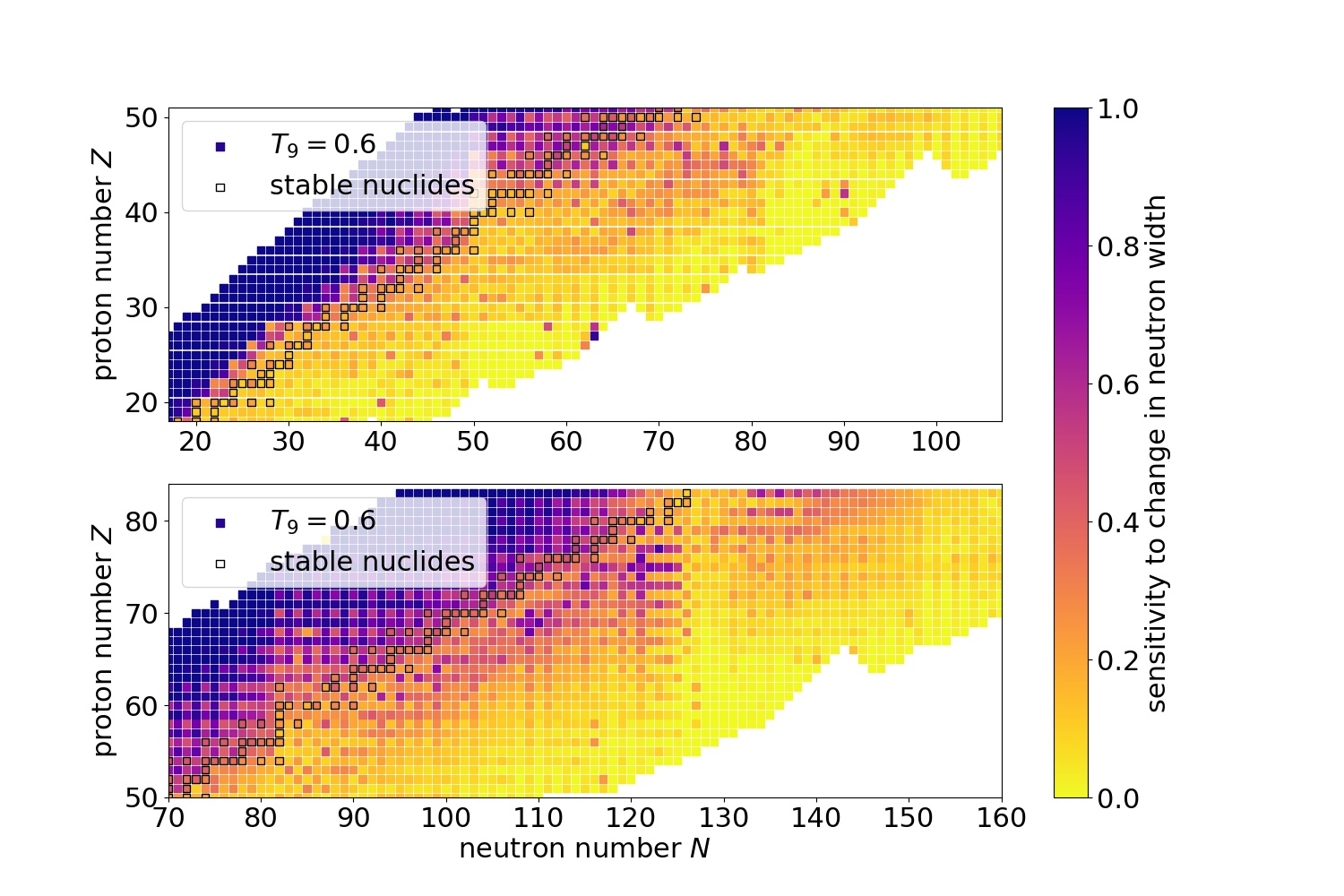}
\caption{Sensitivity of the (n,$\gamma$) rate at 0.6 GK to a change in the neutron width; a value of 1 means that the rate changes equally as the width, a value of 0 means no change. (Data taken from \cite{2012ApJS..201...26R}.)}
\label{fig:sensi}       
\end{figure*}

Even at the low interaction energies of the s-process -- and much more so in high-$T$ processes -- the neutron width is considerably larger than other widths because it is easy to capture or eject uncharged particles with sufficient energy.
Since the cross section (and thus the reaction rate) in the compound nucleus mechanism is determined by the smallest width in entrance or exit channel, (n,$\gamma$) on intermediate and heavy nuclides in astrophysics
are determined by the $\gamma$-width \cite{2012ApJS..201...26R}. This is illustrated in Fig.\ \ref{fig:sensi}. Predictions of the $\gamma$-strength function, however, are notoriously difficult. A further complication is given by the fact that the most important $\gamma$-ray energies
$E_\gamma$, that contribute most to the reaction rate integral, are smaller than the particle emission threshold. This is sketched in Fig.\ \ref{fig:gamma}, where the downward arrows indicate the $\gamma$-emission by de-excitation of the
compound nucleus and the shading indicates the exponential increase of the NLD with increasing excitation energy of nucleus B. As the $\gamma$-strength decreases with decreasing $E_\gamma$ (lengths of the arrows), there is a competition
between decreasing $\gamma$-strength and increasing number of available final states for $\gamma$-decay. This gives rise to a maximum in the emission energy. A realistic example is shown in Fig.\ \ref{fig:maxcontrib} for two Sn isotopes.
Interestingly, calculations across the nuclear chart have shown that the maximum is almost always located about $2-4$ MeV below $\neutronsep$, except very close to the dripline or for nuclides with magic neutron numbers for which the 
compound nucleus model is not applicable \cite{2008PhRvC..78c2801R}.

As mentioned before, the astrophysically relevant neutron energy $E$ is negligible compared to $\neutronsep$ for the majority of applications. This is not correct anymore, however,
when approaching the dripline as in the case of the r-process because of the strongly decreased $\neutronsep$. Nevertheless, the neutron energy $E$ remains low by nuclear physics standards also in the r-process and this
implies that the compound nucleus is formed at low excitation energy. As a consequence, the statistical model may not be applicable anymore and individual resonances and direct capture have to be taken into account. This complicates the
prediction of reaction rates for these nuclides as well as their measurement. It is only consequential, however, in cold r-process scenarios with competition between neutron captures and $\beta^-$-decays whereas in a hot r-process rates
for nuclides close to the dripline do not have to be known \cite{2010ApJ...712.1359F,2011PrPNP..66..346T}. As indicated in Sec.\ \ref{sec:reverse}, an equilibrium is established in a hot r-process and the calculation of the equilibrium abundances only requires the knowledge of $\neutronsep$
along with $T$ and $n_\rmn$ \cite{Rauscherb}.


\section{Summary and conclusions}
\label{sec:summary}

It is important to keep in mind that methods developed to determine reaction rates for low-$T$ nucleosynthesis may not be applicable to processes at high $T$. This is because reactions on nuclei in excited states dominate the
astrophysical reaction rates, involving many more transitions than reactions on nuclei in their ground states. This also limits the usefulness of the MACS, which was developed for a direct experimental determination of neutron capture rates
for the s-process,
especially because it was found that even at s-process temperatures thermally excited states contribute more to the rate than previously assumed. Moreover, different nuclear properties (such as resonance energies and widths) may be of higher or lower importance
depending on the reaction mechanism. Going to even higher temperature and higher mass number of the involved nuclides,
these circumstances conspire to simplify a theoretical treatment (with exception of reactions on magic nuclei and close to the driplines) and complicate an experimental constraint of astrophysical rates. The many transitions (from target states to final states, mostly via compound states) involved in explosive nucleosynthesis lend themselves to the use of averaged quantities in their prediction. On the other hand, the large number of transitions restricts the applicability of direct and indirect experimental approaches studying a few transitions, as usually applied in the study of light nuclei, even when dealing with stable nuclides.

In experimental investigations, it has to be made sure that astrophysically relevant properties are studied and these may be different at low and at high temperature because the reaction mechanism may be changing. The cited systematic sensitivity studies and g.s.\ contributions to the astrophysical rate can help to guide experiments. Regarding theory, predictions are simplified for high-$T$ rates by being able to average over many transitions and apply the Hauser-Feshbach model, which has been successful in describing a large number of reaction cross sections. Nevertheless, considerable challenges for nuclear theory remain. First, nuclear structure models have to be improved to be able to reliably predict the nuclear properties (such as nuclear spectroscopy, NLD, or $\gamma$-strength functions) required for the astrophysical reaction cross section prediction. Another major challenge to theory is to accurately describe the competition between direct, resonant, and statistical reaction mechanisms for nuclei close to magic numbers and close to the driplines. Some first attempts have been made to combine direct and Hauser-Feshbach cross sections for neutron-rich nuclei \cite{rauintjmodphys,2014PhRvC..90b4604X} but currently a reliable prediction of individual resonances and their interference (which nevertheless may be very important also for magic nuclei and close to driplines) is beyond the reach of theory.


\bibliographystyle{spphys}       
\bibliography{rauscher.bib}   

\begin{thebibliography}{10}
\providecommand{\url}[1]{{#1}}
\providecommand{\urlprefix}{URL }
\expandafter\ifx\csname urlstyle\endcsname\relax
  \providecommand{\doi}[1]{DOI \discretionary{}{}{}#1}\else
  \providecommand{\doi}{DOI \discretionary{}{}{}\begingroup
  \urlstyle{rm}\Url}\fi

\bibitem{B2FH}
E.M. Burbidge, G.R. Burbidge, W.A. Fowler, F.~Hoyle, \rmp \textbf{29}, 547
  (1957)

\bibitem{Cameron57}
A.G.W. {Cameron}, Publ. Astron. Soc. Pac. \textbf{69}, 201 (1957)

\bibitem{1977ApJ...212..149C}
J.J. {Cowan}, W.K. {Rose}, \apj \textbf{212}, 149 (1977).
\newblock \doi{10.1086/155030}

\bibitem{1978ApJS...36..285W}
S.E. {Woosley}, W.M. {Howard}, \apjs \textbf{36}, 285 (1978).
\newblock \doi{10.1086/190501}

\bibitem{Rauscher2}
T.~Rauscher, N.~Dauphas, I.~Dillmann, C.~Fr\"ohlich, Z.~F\"ul\"op, G.~Gy\"urky,
  Rep. Prog. Phys. \textbf{76}, 066201 (2013)

\bibitem{rauintjmodphys}
T.~{Rauscher}, International Journal of Modern Physics E \textbf{20}(5), 1071
  (2011).
\newblock \doi{10.1142/S021830131101840X}

\bibitem{Rauscherb}
T.~Rauscher, \emph{Essentials of Nucleosynthesis and Theoretical Nuclear
  Astrophysics} (IOP Publishing, Bristol, 2020)

\bibitem{1969ApJS...18..247W}
R.V. {Wagoner}, The Astrophysical Journal Supplement \textbf{18}, 247 (1969).
\newblock \doi{10.1086/190191}

\bibitem{rauscher3}
T.~{Rauscher}, Physical Review C \textbf{81}, 045807 (2010)

\bibitem{1961AnPhy..12..331C}
D.D. {Clayton}, W.A. {Fowler}, T.E. {Hull}, B.A. {Zimmerman}, Annals of Physics
  \textbf{12}(3), 331 (1961).
\newblock \doi{10.1016/0003-4916(61)90067-7}

\bibitem{1965ApJS...11..121S}
P.A. {Seeger}, W.A. {Fowler}, D.D. {Clayton}, \apjs \textbf{11}, 121 (1965).
\newblock \doi{10.1086/190111}

\bibitem{1979LIACo..22...79H}
L.D. {Hong}, H.~{Beer}, F.~{Kaeppeler}, in \emph{Liege International
  Astrophysical Colloquia}, \emph{Liege International Astrophysical Colloquia},
  vol.~22, ed. by A.~{Boury}, N.~{Grevesse}, L.~{Remy-Battiau} (1979),
  \emph{Liege International Astrophysical Colloquia}, vol.~22, pp. 79--93

\bibitem{1980PhRvC..21..534B}
H.~{Beer}, F.~{K{\"a}ppeler}, \prc \textbf{21}(2), 534 (1980).
\newblock \doi{10.1103/PhysRevC.21.534}

\bibitem{1982ApJ...257..821K}
F.~{Kaeppeler}, H.~{Beer}, K.~{Wisshak}, D.D. {Clayton}, R.L. {Macklin}, R.A.
  {Ward}, \apj \textbf{257}, 821 (1982).
\newblock \doi{10.1086/160033}

\bibitem{2014JPhG...41e3101R}
R.~{Reifarth}, C.~{Lederer}, F.~{K{\"a}ppeler}, Journal of Physics G Nuclear
  Physics \textbf{41}(5), 053101 (2014).
\newblock \doi{10.1088/0954-3899/41/5/053101}

\bibitem{2000ADNDT..76...70B}
Z.Y. {Bao}, H.~{Beer}, F.~{K{\"a}ppeler}, F.~{Voss}, K.~{Wisshak},
  T.~{Rauscher}, Atomic Data and Nuclear Data Tables \textbf{76}(1), 70 (2000).
\newblock \doi{10.1006/adnd.2000.0838}

\bibitem{2006AIPC..819..123D}
I.~{Dillmann}, M.~{Heil}, F.~{K{\"a}ppeler}, R.~{Plag}, T.~{Rauscher}, F.K.
  {Thielemann}, in \emph{Capture Gamma-Ray Spectroscopy and Related Topics},
  \emph{American Institute of Physics Conference Series}, vol. 819, ed. by
  A.~{Woehr}, A.~{Aprahamian} (2006), \emph{American Institute of Physics
  Conference Series}, vol. 819, pp. 123--127.
\newblock \doi{10.1063/1.2187846}

\bibitem{2014NDS...120..171D}
I.~{Dillmann}, T.~{Sz{\"u}cs}, R.~{Plag}, Z.~{F{\"u}l{\"o}p},
  F.~{K{\"a}ppeler}, A.~{Mengoni}, T.~{Rauscher}, Nuclear Data Sheets
  \textbf{120}, 171 (2014).
\newblock \doi{10.1016/j.nds.2014.07.038}

\bibitem{1999ApJ...525..886A}
C.~{Arlandini}, F.~{K{\"a}ppeler}, K.~{Wisshak}, R.~{Gallino}, M.~{Lugaro},
  M.~{Busso}, O.~{Straniero}, \apj \textbf{525}(2), 886 (1999).
\newblock \doi{10.1086/307938}

\bibitem{1980ApJ...238..266W}
R.A. {Ward}, W.A. {Fowler}, \apj \textbf{238}, 266 (1980).
\newblock \doi{10.1086/157983}

\bibitem{1974QJRAS..15...82F}
W.A. {Fowler}, \qjras \textbf{15}, 82 (1974)

\bibitem{2011ApJ...738..143R}
T.~{Rauscher}, P.~{Mohr}, I.~{Dillmann}, R.~{Plag}, \apj \textbf{738}(2), 143
  (2011).
\newblock \doi{10.1088/0004-637X/738/2/143}

\bibitem{2012ApJ...755L..10R}
T.~{Rauscher}, Ap. J. Lett. \textbf{755}(1), L10 (2012).
\newblock \doi{10.1088/2041-8205/755/1/L10}

\bibitem{2012ApJS..201...26R}
T.~{Rauscher}, \apjs \textbf{201}(2), 26 (2012).
\newblock \doi{10.1088/0067-0049/201/2/26}

\bibitem{2008PhRvL.101s1101K}
G.G. {Kiss}, T.~{Rauscher}, G.~{Gy{\"u}rky}, A.~{Simon}, Z.~{F{\"u}l{\"o}p},
  E.~{Somorjai}, \prl \textbf{101}(19), 191101 (2008).
\newblock \doi{10.1103/PhysRevLett.101.191101}

\bibitem{2009PhRvC..80c5801R}
T.~{Rauscher}, G.G. {Kiss}, G.~{Gy{\"u}rky}, A.~{Simon}, Z.~{F{\"u}l{\"o}p},
  E.~{Somorjai}, \prc \textbf{80}(3), 035801 (2009).
\newblock \doi{10.1103/PhysRevC.80.035801}

\bibitem{2005NuPhA.758..655R}
T.~{Rauscher}, \nphysa \textbf{758}, 655 (2005).
\newblock \doi{10.1016/j.nuclphysa.2005.05.160}

\bibitem{2006NuPhA.777..137D}
P.~{Descouvemont}, T.~{Rauscher}, \nphysa \textbf{777}, 137 (2006).
\newblock \doi{10.1016/j.nuclphysa.2004.10.024}

\bibitem{2018arXiv180310581R}
T.~{Rauscher}, arXiv e-prints arXiv:1803.10581 (2018)

\bibitem{2008PhRvC..78c2801R}
T.~{Rauscher}, \prc \textbf{78}(3), 032801 (2008).
\newblock \doi{10.1103/PhysRevC.78.032801}

\bibitem{2010ApJ...712.1359F}
K.~{Farouqi}, K.L. {Kratz}, B.~{Pfeiffer}, T.~{Rauscher}, F.K. {Thielemann},
  J.W. {Truran}, \apj \textbf{712}(2), 1359 (2010).
\newblock \doi{10.1088/0004-637X/712/2/1359}

\bibitem{2011PrPNP..66..346T}
F.K. {Thielemann}, A.~{Arcones}, R.~{K{\"a}ppeli}, M.~{Liebend{\"o}rfer},
  T.~{Rauscher}, C.~{Winteler}, C.~{Fr{\"o}hlich}, I.~{Dillmann}, T.~{Fischer},
  G.~{Martinez-Pinedo}, K.~{Langanke}, K.~{Farouqi}, K.L. {Kratz}, I.~{Panov},
  I.K. {Korneev}, Progress in Particle and Nuclear Physics \textbf{66}(2), 346
  (2011).
\newblock \doi{10.1016/j.ppnp.2011.01.032}

\bibitem{2014PhRvC..90b4604X}
Y.~{Xu}, S.~{Goriely}, A.J. {Koning}, S.~{Hilaire}, \prc \textbf{90}(2), 024604
  (2014).
\newblock \doi{10.1103/PhysRevC.90.024604}

\end{thebibliography}


\end{document}